# AI-assisted writing and the reorganization of scientific knowledge

Erjia Yan[1*], Chaoqun Ni[2]

[1]College of Computing and Informatics, Drexel University, 3141 Chestnut Street, Philadelphia, PA 19104

[2] School of Computer, Data, and Information Sciences (CDIS), University of Wisconsin-Madison, 1205 University Ave, Madison, WI 53706

[*]Corresponding author: ey86@drexel.edu

## Significance statement

Generative AI is rapidly entering scientific writing, but its effects on how science builds upon prior knowledge remain unclear. We analyze 2 million full-text research articles linked to citation networks and show that, after the diffusion of large language models in 2023, AI-assisted writing intensity became positively associated with citation disruption. At the same time, its association with broad cross-field citation sourcing weakened, indicating more disruptive work built from relatively narrower knowledge inputs. These findings suggest that AI may reshape scientific discovery not only by increasing the rate of writing but also by altering how prior knowledge is combined and displaced.

## ABSTRACT

Generative AI systems such as ChatGPT are increasingly used in scientific writing, yet their broader implications for the organization of scientific knowledge remain unclear. We examine whether AI-assisted writing intensity, measured as the share of text in a paper that is predicted to exhibit features consistent with LLM–generated text, is associated with scientific disruption and knowledge recombination. Using approximately two million full-text research articles published between 2021 and 2024 and linked to citation networks, we document a sharp temporal pattern beginning in 2023. Before 2023, higher AI-assisted writing intensity is weakly or negatively associated with disruption; after 2023, the association becomes positive in within-author, within-field analyses. Over the same period, the positive association between AI-assisted writing intensity and cross-field citation breadth weakens substantially, and the negative association with citation concentration attenuates. Thus, the post-2023 increase in disruption is not accompanied by broader knowledge sourcing. These patterns suggest that generative AI is associated with more disruptive citation structures without a corresponding expansion in cross-field recombination. Rather than simply broadening the search space of science, AI-assisted writing may be associated with new forms of recombination built from relatively narrower knowledge inputs.

Generative artificial intelligence is rapidly changing how scientific work is written and communicated (1-3). Large language models (LLMs) can generate fluent text, summarize literature, suggest references, and assist with drafting and revision (4, 5). Their growing use in research raises a broader question: do these tools simply lower the cost of scientific communication, or do they also alter how scientific knowledge is combined and displaced?

Historically, new communication technologies have reshaped the production and circulation of knowledge (6, 7). Such changes can widen intellectual exploration by reducing barriers to access and information synthesis, but they can also reinforce existing hierarchies by steering attention toward already salient ideas and sources (8). Generative AI raises both possibilities; because these systems are trained on large corpora of existing text, they may favor canonical literatures and familiar argumentative patterns. But by making synthesis faster and cheaper, they may also facilitate combinations of ideas that would otherwise be more difficult to assemble.

One way to study this question is through citation networks. The disruption index captures whether a focal paper tends to displace antecedent work; that is, to be cited without its references, or instead to consolidate existing knowledge by being cited together with earlier work (9, 10). Prior research shows that highly disruptive contributions are rare and that science has become increasingly consolidative over time (11, 12). If generative AI is changing the organization of scientific knowledge, those changes should be visible in the citation structures surrounding papers.

We study this question using approximately two million full-text research articles published between 2021 and 2024. For each paper, we measure AI-assisted writing intensity using the classifier-based indicator developed in (13), linking the paper to its citation network, and compute disruption, cross-field citation breadth, and citation concentration. Because our empirical design compares the same author working in the same field over time, the analysis is identified from within-author, within-field variation rather than cross-sectional differences across researchers or disciplines.

Our analysis identifies a sharp reversal beginning in 2023. Before 2023, higher AI-assisted writing intensity is weakly or negatively associated with disruption; after 2023, the association becomes positive. Over the same period, the association between AI-writing intensity and broad cross-field sourcing weakens substantially. These results suggest that generative AI is associated with a reconfiguration of citation structure: papers with higher AI-writing intensity become more disruptive, but not because they draw on a broader set of fields. Instead, the evidence is more consistent with changing patterns of recombination within relatively narrower knowledge inputs.

## RESULTS

### Overview

We examine whether AI-assisted writing intensity is associated with citation behavior, and whether this relationship shifted following the widespread diffusion of large language models in 2023. AI-assisted writing intensity is measured as the percentage of text in a paper that was likely generated by LLMs (13). As a probabilistic measure, it takes non-zero values prior to 2023, reflecting background variation in standardized writing and earlier language technologies. Because the analysis is conducted at the author–field–year level and identified using within-author variation over time, time-invariant measurement error is differenced out. Identification therefore relies on the discrete change in the relationship between this measure and citation outcomes following the diffusion of generative AI. We combine a structural break design with within-author and within-field identification strategies that eliminate compositional bias (14, 15). Specifically, we first test time based breaks and delineations in the AI–disruption relationship, then assess whether the observed patterns reflect within-author behavioral change.

**The AI–disruption association reverses after 2023**

We begin by estimating year-specific associations between AI-assisted writing intensity and the consolidation–disruption (CD) index (10-12, 16), absorbing author×field and year fixed effects. This specification ensures that identification relies exclusively on changes within the works of the same author in the same field over time (14, 15).

The relationship between AI-assisted writing intensity and scientific disruption exhibits a sharp temporal pattern (Figure 1). In 2021, higher values of the AI-assisted writing intensity measure (reflecting AI-like linguistic features rather than direct tool usage) are associated with more consolidative citation patterns, with negative marginal effects ranging from −0.095 to −0.227 across citation windows. By 2022, this relationship converges toward zero. Beginning in 2023, however, the association becomes strongly positive and remains so in 2024, indicating that papers with higher levels of AI-assisted writing are increasingly associated with more disruptive citation structures.

The reversal is highly consistent across 12-, 18-, and 24-month citation windows, which argues against citation-window artifacts. Because identification comes from within-author changes within fields, the pattern is more consistent with behavioral adjustment than with static compositional differences across researchers or research areas.

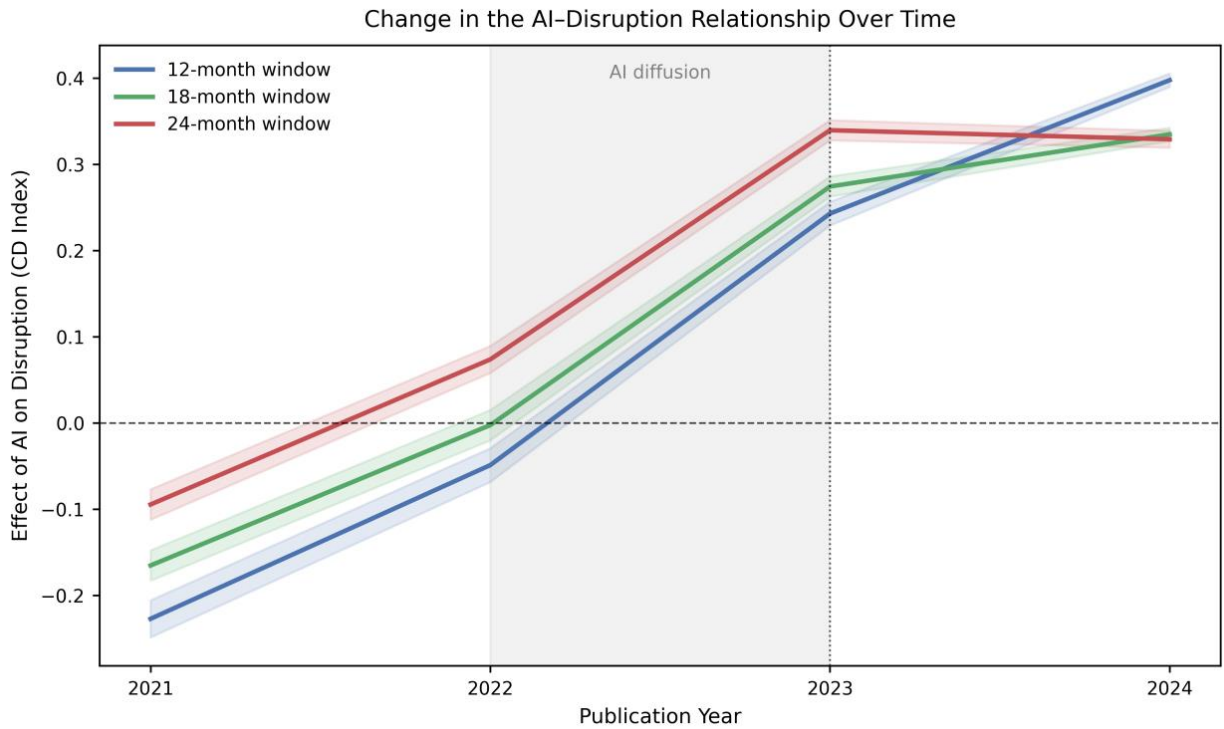

**Figure 1.** Year-specific associations between AI-assisted writing intensity and scientific disruption. Marginal effects of AI-assisted writing intensity on the consolidation–disruption (CD) index are estimated from author–field–year panel models with author×field and year fixed effects. Shaded bands show 95% confidence intervals. In 2021, the association is negative across all forward-citation windows; in 2022, it attenuates toward zero; in 2023 and 2024, it becomes positive and remains so. The dashed line marks 2023, the first full year after the public diffusion of advanced generative AI tools.

**Within-author change models are consistent with behavioral adjustment**

To distinguish within-person adjustment from changing sample composition, we next estimate first-difference models that relate changes in AI-assisted writing intensity to contemporaneous changes in disruption within the same author–field unit. This specification asks whether authors whose AI-assisted writing intensity rises more over time also exhibit larger increases in disruption.

In the baseline 18-month specification, a one-unit increase in AI-assisted writing intensity is associated with a 0.151 increase in the CD index (β = 0.151, SE = 0.004, $p <$ 0.001). The magnitude is highly consistent across alternative windows (12m: β = 0.152, SE = 0.005; 24m: β = 0.171, SE = 0.005; all $p <$ 0.001), indicating that the result is not driven by citation windows or truncation effects.

These estimates imply that even moderate increases in AI-assisted writing intensity correspond to meaningful shifts in citation structure. Because identification relies entirely on within-author variation, these findings indicate that the relationship between AI-assisted writing and disruption reflects behavioral adjustment rather than the selection of inherently more disruptive researchers into AI use.

**The association with broad cross-field sourcing weakens after 2023**

We next examine whether AI-assisted writing intensity is associated with the structure of knowledge inputs. Before 2023, higher AI-assisted writing intensity is strongly associated with broader cross-field citation breadth and lower citation concentration. In the 18-month window, the estimated effect on entropy is 1.277 in 2021 and 1.225 in 2022, whereas the estimated effect on HHI is −0.561 in 2021 and −0.539 in 2022. After 2023, both associations attenuate substantially. By 2024, the estimated effect on entropy remains positive but falls to 0.455, and the estimated effect on HHI remains negative but moves closer to zero at −0.207. The same pattern appears in the 12- and 24-month windows.

These estimates do not show that papers with higher AI-assisted writing intensity become narrower or more concentrated in absolute terms after 2023. Rather, they show that the earlier association between AI-assisted writing intensity and broader, more diffuse knowledge sourcing becomes much weaker. The post-2023 increase in disruption is therefore not accompanied by a corresponding expansion in cross-field recombination. A cautious interpretation is that generative AI becomes associated with more disruptive citation outcomes even as its link to broad knowledge inputs weakens.

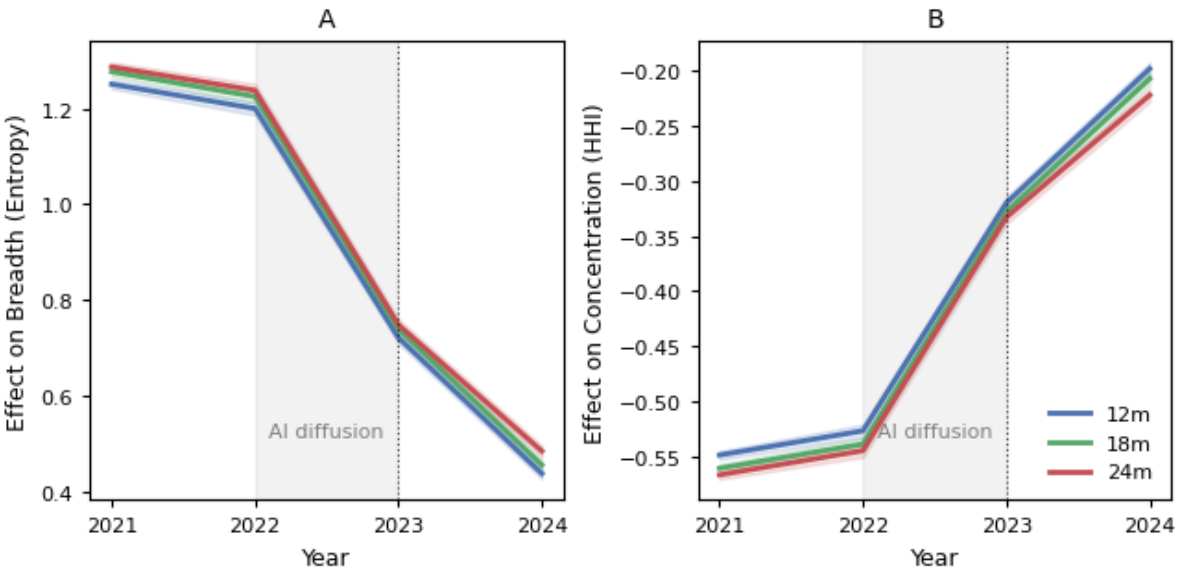

**Figure 2. The association between AI-assisted writing intensity and broad knowledge sourcing weakens after 2023.** (A) Year-specific marginal effects of AI-assisted writing intensity on cross-field citation breadth (entropy) and (B) citation concentration (HHI), estimated from author–field–year panel models with author×field and year fixed effects. Shaded bands show 95% confidence intervals. Before 2023, higher AI-assisted writing intensity is strongly associated with broader cross-field sourcing and lower concentration. After 2023, both associations remain in the same direction but attenuate substantially.

**Robustness and heterogeneity**

The results are robust across a range of alternative specifications. The structural reversal in the AI–disruption relationship is consistent across citation windows (12-month, 18-month, and 24-month), ruling out citation window artifacts (Table S1), and is confirmed by within-author acceleration models, which link increases in AI use to contemporaneous increases in disruption (Table S2). Analyses of citation structure show that the diffusion of generative AI is associated with reduced cross-field recombination and more concentrated knowledge inputs, as reflected in declines of entropy and dispersion (Tables S3–S4).

To address potential concerns about authorship assignment, we re-estimate the model using only first-author observations, which more directly capture manuscript-level writing behavior. The results remain qualitatively identical but exhibit substantially larger magnitudes: the association is strongly negative in 2021 ($\beta_{(2021)} = -0.395$, SE =

0.024), attenuates in 2022 ($\beta_{(2022)}$ = −0.116, SE = 0.022), and becomes strongly positive in 2023–2024 ($\beta$ = 0.321–0.408, SE = 0.014–0.008). This increase relative to the full sample is consistent with attenuation arising from assigning AI exposure to all coauthors.

Inference is stable across variance estimators, with clustered standard errors yielding identical coefficients (Table S5), and results are unchanged when controlling for team size (Table S6). Additional analyses reveal modest heterogeneity by researcher seniority and stronger effects in fields with more diverse knowledge structures, while differences across linguistic and geographic contexts are small and statistically insignificant in the post-2023 period (Sections S8–S9). Placebo tests using the pre-period show no evidence of a structural shift prior to 2023 ($\beta_{(2023)}$ = 0.001, $p$ = 0.91), confirming that the observed reversal is not driven by pre-existing trends (Section S13).

## DISCUSSION

Our findings identify a temporal pattern in the association between AI-assisted writing and the structure of scientific citation networks. Before 2023, higher AI-assisted writing intensity is weakly or negatively associated with disruption; beginning in 2023, the association becomes positive in within-author analyses. This shift coincides with the rapid diffusion of advanced generative AI tools into scientific workflows.

This pattern is notable within the context of evidence that scientific progress has become increasingly consolidative (11). Against this backdrop, the observed shift suggests that generative AI is associated with a departure from recent trends toward incrementalism. It suggests that generative AI may be associated with a partial departure from recent trends toward incrementalism. At the same time, the results do not imply that AI mechanically causes novel discovery. They show a change in citation structure associated with higher AI-assisted writing intensity, not a direct measure of substantive creativity or scientific value.

The most important qualification is that the post-2023 increase in disruption is not paired with broader cross-field sourcing. Before 2023, higher AI-assisted writing intensity is strongly associated with broader citation breadth and lower concentration. After 2023, those relationships remain in the same direction but become much weaker. Thus, papers with higher AI-assisted writing intensity continue to draw on somewhat broader and less concentrated references than lower-AI papers, but far less so than in the pre-diffusion period.

One interpretation is that generative AI lowers the cost of synthesizing and reorganizing literature that is already close to a researcher's focal domain. Under that interpretation, AI assistance need not broaden the set of fields a paper draws from; instead, it may facilitate new combinations among more proximate sources within a field. Our evidence is consistent with that possibility. In this sense, AI-assisted writing may facilitate a form of recombination that is structurally distinct from traditional interdisciplinary integration.

Several limitations follow. First, the analysis is observational and cannot establish a causal effect of AI-assisted writing on scientific outcomes. Unobserved changes in journals, fields, editorial norms, or research topics may also contribute to the observed pattern. Second, our measure of AI-assisted writing intensity is a classifier-based proxy and should not be interpreted as a direct record of tool use. Third, the disruption index captures structural relationships in citation networks rather than the substantive content or long-run importance of ideas.

Even with these caveats, the results point to a meaningful change in how scientific work is organized in citation space. AI-assisted writing is associated with more disruptive citation structures after 2023, but not with a parallel broadening of knowledge inputs across fields. If these patterns persist, generative AI may alter the balance between consolidation and displacement in science by reshaping how prior knowledge is recombined rather than simply by accelerating writing.

**MATERIALS AND METHODS**

**Data construction**

For this paper, we analyze full-text research articles from PubMed Central published between 2021 and 2024 that are linked to OpenAlex metadata and citation records(13). The analytic dataset comprises approximately two million works, which we aggregate into author–field–year panels to enable author identification over time.

We construct outcomes using three forward-citation windows—12, 18, and 24 months—defined relative to each paper's publication date. For a focal work published on date t, we count only forward citations appearing within the corresponding window after t. Because the OpenAlex snapshot used in this study was updated on March 6, 2026, the 18-month specification includes papers published on or before September 6, 2024, and the 24-month specification includes papers published on or before March 6, 2024. For the year-interacted models, we retain author–field units observed in more than one author×year cell.

The resulting panels contain 5,437,536 observations in the 12-month window, 5,634,835 observations in the 18-month window, and 5,124,693 observations in the 24-month window. Aggregation to the author–field–year level reduces paper-level idiosyncratic noise while preserving within-author variation across time and disciplinary context.

Fixed forward-citation windows ensure comparability across publication cohorts. Without such windows, more recent papers would mechanically appear less cited or less disruptive simply because they have had less time to accumulate forward citations. Restricting attention to equal exposure windows reduces truncation bias and makes estimates comparable across years.

**AI-assisted writing measure**

For each work, we construct a continuous measure of AI-assisted writing intensity, defined as the share of text in a paper that is predicted to exhibit features consistent with large language model (LLM)–generated text (13). The underlying estimator is based on a validated distributional framework that models each document as a mixture of human- and AI-generated sentence distributions, achieving low estimation error (≈3.5%) and showing consistency with independent AI-detection signals while avoiding the high false-positive rates of classifier-based approaches. Because this measure captures probabilistic similarity rather than direct tool usage, it takes non-zero values prior to 2023, reflecting background variation in standardized academic writing and earlier language technologies (e.g., grammar correction, predictive text, and early LLM-based tools such as GPT-3).

A central concern is that the measure could reflect stylistic convergence rather than actual AI assistance. Several features of the design and results mitigate this concern. First, the measure relies on full lexical distributions rather than keyword or stylistic markers, reducing sensitivity to superficial changes in academic tone. Second, the analysis is conducted at the author–field–year level and identified using within-author variation over time, differencing out time-invariant writing style and baseline measurement error. Third, the empirical strategy does not rely on absolute levels of the measure, but on the discrete post-2022 shift following the diffusion of generative AI. Consistent with a technology adoption interpretation, this shift is sharp, temporally aligned with ChatGPT, and exhibits systematic heterogeneity: being larger in lower English-proficiency contexts and among authors with prior AI experience, patterns that are difficult to reconcile with uniform changes in writing norms. We therefore interpret the measure as capturing text-level AI assistance, while acknowledging that residual stylistic confounding and under-detection of heavily revised or non-textual uses cannot be fully excluded.

**Metrics**

Our primary outcome is the simplified consolidation–disruption (CD) index, a forward-citation–based measure of structural displacement in the citation network(12, 16). The simplified CD index is defined as:

$$CD = \frac{N_i - N_j}{N_i + N_j}$$

where $N_i$ counts forward citations to the focal paper that do not cite its references, and $N_j$ counts forward citations that co-cite both the focal paper and its references(12, 16). Higher values indicate greater displacement of prior knowledge.

The simplified version compares forward citations that cite the focal work but not its references to those that co-cite both the focal work and its references. We adopt the simplified formulation because the simplified CD index preserves the core logic of citation displacement while allowing computation under author×field panel models. Results are robust to the full CD formulation (Section S11).

To examine mechanisms, we construct two complementary measures of knowledge recombination based on reference fields. First, we compute citation breadth (entropy) as(17):

$$H = -\sum_{f} p_f \log(p_f)$$

where $p_f$ denotes the share of references in field $f$ (n = 26). Second, we compute citation concentration (Herfindahl index, HHI) as(18):

$$HHI = \sum_{f} p_f^2$$

Higher entropy values indicate broader cross-field combinations, whereas higher HHI indicates greater concentration of citations within a smaller set of fields. These measures provide complementary views of how knowledge inputs are distributed.

**Modeling strategy**

Our empirical strategy is designed to isolate an author's behavioral changes and distinguish structural composition shifts from compositional differences across researchers.

First, we estimate fixed-effects panel models with author×field and year fixed effects, ensuring that identification relies exclusively on within-author changes over time. This allows comparisons of the same researcher operating in the same disciplinary context before and after the diffusion of large language models.

Second, we allow the relationship between AI-assisted writing intensity and outcomes to vary flexibly by year. This specification enables direct detection of temporal shifts rather than imposing a constant association.

Third, to test whether observed patterns reflect behavioral change, we estimate within-author acceleration models relating changes in AI-assisted writing intensity to contemporaneous changes in disruption. This dynamic specification distinguishes behavioral adjustment from static sorting.

Fourth, we examine changes in the structure of knowledge inputs by estimating analogous models for entropy and HHI. These models allow us to assess whether AI-assisted writing is associated with broader recombination across fields or with increased concentration of references.

Finally, we assess heterogeneity by academic age and the linguistic context (English-speaking vs. non-English-speaking) by using continuous interaction models. Academic age is defined as years since first observed publication and capped at 50 years to address metadata outliers; observations at the cap are excluded to avoid distortion from top-coding. This specification allows the effect of AI-assisted writing to vary smoothly across career stages. The linguistic context is determined by a broad definition in which a country is considered English-speaking if more than 20% of its people speak English (13) and a strict definition which includes several English-speaking countries (United

States, United Kingdom, Canada, Australia, New Zealand, Ireland, Singapore, and South Africa).

**Estimating year-specific shifts in the AI–outcome association**

Our central empirical question is whether the relationship between AI-assisted writing and citation disruption changes after the widespread diffusion of large language models beginning in late 2022. A simple cross-sectional comparison would confound AI-assisted writing with stable differences across authors or fields. We therefore rely on panel models with fixed effects (14, 19).

To detect a structural discontinuity, we estimate a year-interacted model:

$$Y_{iaft} = \sum_t \beta_t (AI_{iaft} \cdot I_t) + \delta_{a \times f} + \lambda_t + \varepsilon_{iaft}$$

where $\delta_{a \times f}$ are author×field fixed effects and $\lambda_t$ are year fixed effects and $Y$ represents disruption (CD), entropy, or HHI.

The inclusion of author×field fixed effects ensures that identification comes from within-author changes over time while holding disciplinary specialization constant; year fixed effects absorb aggregate shocks common to all scholars. The interaction terms $\beta_t$ allow the association between AI-assisted writing intensity and each outcome to vary across years, enabling detection of temporal shifts that coincide with the diffusion of generative AI. Estimation is implemented via Frisch–Waugh–Lovell (FWL) residualization (20), yielding a two-way fixed effects specification in which both the dependent variable and regressors are orthogonalized with respect to the author×field and year effects prior to coefficient estimation.

**Within-author change model**

Even a structural break could reflect sorting rather than behavioral change. To isolate dynamic adjustments within individuals, we estimate an acceleration model that regresses disruption on within-author changes in AI-assisted writing intensity(21):

$$CD_{iaft} = \beta_t \Delta AI_{iaft} + \delta_{a \times f} + \lambda_t + \varepsilon_{iaft}$$

where changes are computed within author×field units. This specification tests whether increases in AI-assisted writing intensity within the same researcher are associated with contemporaneous changes in outcomes.

**Inference and robustness**

All models use heteroskedasticity-robust (HC1) standard errors, and key specifications are re-estimated with standard errors clustered at the author–field level. Fixed effects are implemented via within transformation.

Robustness checks include alternative citation windows, first-difference models, first-author restrictions, placebo tests, alternative variance estimators, team-size controls, and the full CD formulation. Together, these checks evaluate whether the main pattern is sensitive to citation dynamics, inference choices, authorship assignment, or model specification.

## Competing interest statement

The authors declare no competing interests.

## Author contributions

EY: Conceptualization, Data curation, Methodology, Visualization, Writing-Original draft preparation. CN: Investigation, Writing-Original draft preparation.

## AI use statement

Portions of code development and text editing were assisted using a large language model (GPT-5.3). All outputs were reviewed, validated, and edited by the authors.

## Data and code availability

All data needed to evaluate the conclusions in the paper are present in the paper and/or the Supplementary Materials. The datasets generated and analyzed during the current study are available in Zenodo at https://doi.org/10.5281/zenodo.19581823. Code used to conduct event study analysis is available in GitHub at

https://github.com/erjiayan/AIdisruptiveness

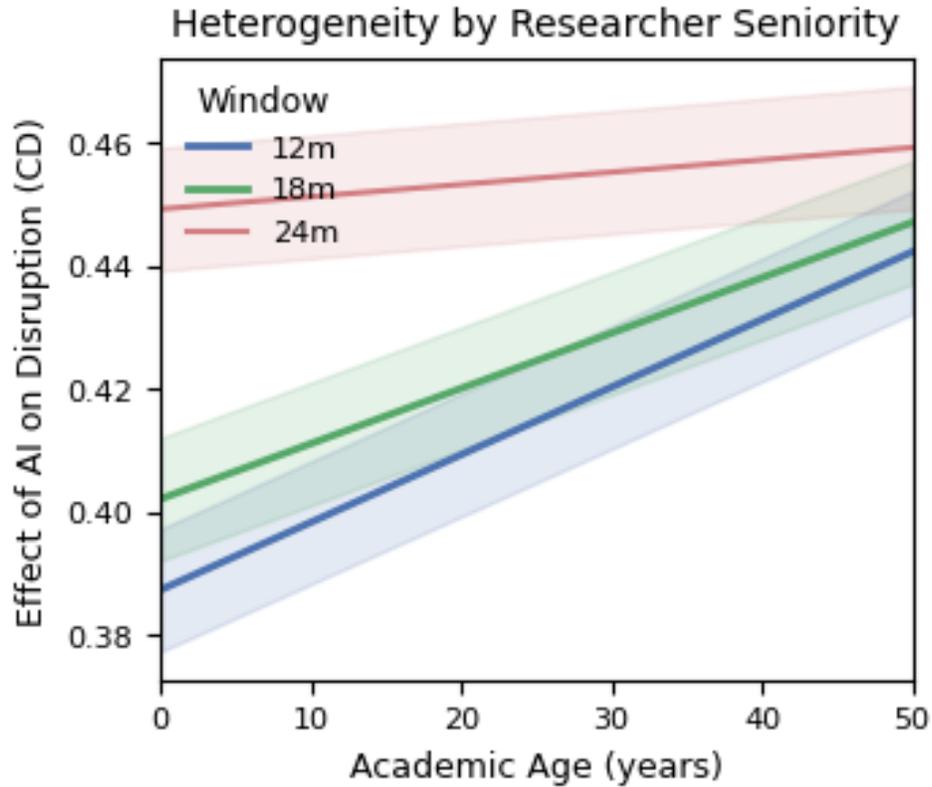

**Fig. S1. Heterogeneity in the effect of AI-assisted writing intensity by researcher seniority.** The interaction between AI-assisted writing intensity and academic age is positive and statistically significant in the 12- and 18-month windows (12m: β = 0.0011, 18m: β = 0.0009; both $p < 0.001$) but small and statistically indistinguishable from zero in the 24-month window (β = 0.0002, $p = 0.48$). Academic age is capped at 50 years, and observations at the cap are excluded from the analysis. The pattern indicates that any seniority-related advantage is modest and concentrated in shorter citation windows.

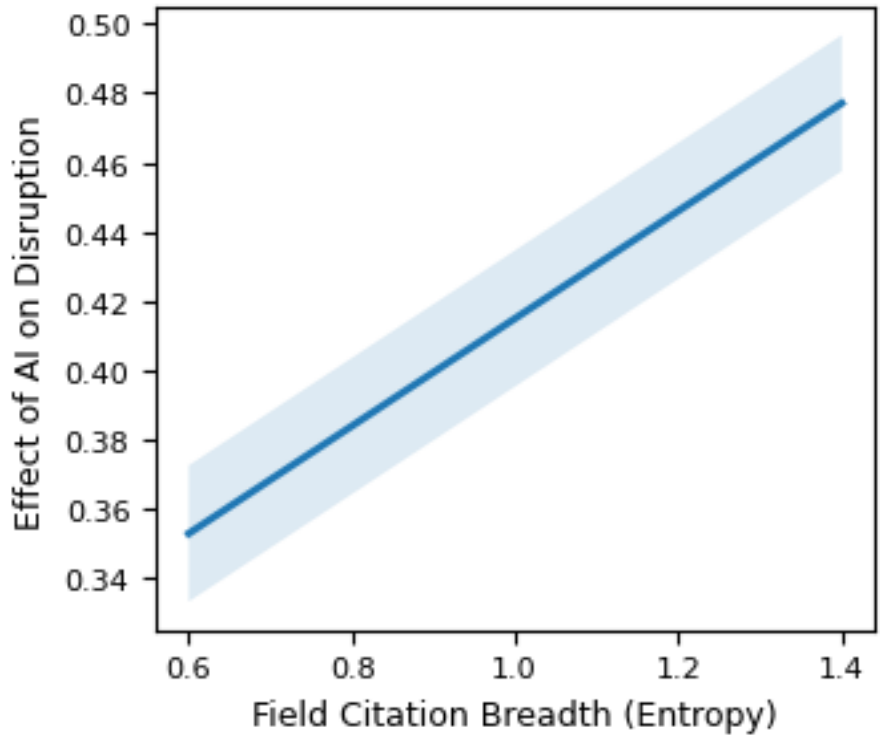

**Fig. S2. Field-level heterogeneity in the AI–disruption relationship.** The marginal effect of AI-assisted writing intensity on disruption increases with baseline field-level citation breadth using the 18m citation window. In fields with more diverse knowledge structures (higher entropy), AI is associated with a larger increase in disruption. Shaded areas represent approximate 95% confidence intervals. The association between AI-assisted writing intensity and disruption varies systematically across fields. In fields with greater baseline citation breadth, the effect of AI on disruption is significantly stronger ($\beta = 0.155$, SE = 0.011, $p < 0.001$). This suggests that the impact of AI depends on the underlying structure of knowledge within a field, with larger effects observed in more interdisciplinary domains.

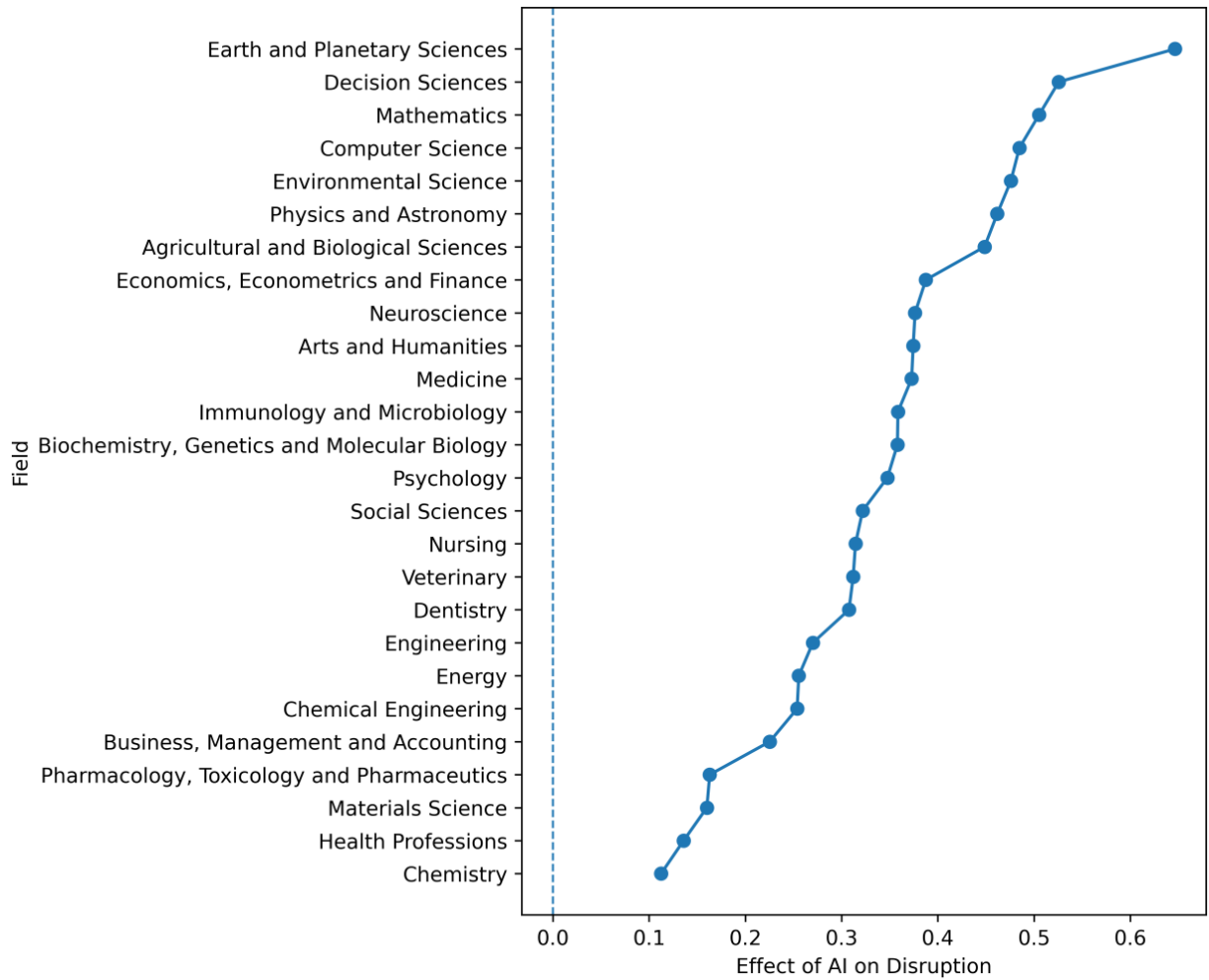

**Fig. S3. Field-level variation in the AI–disruption relationship.** Estimated marginal effects of AI-assisted writing intensity on disruption (CD index) across 26 scientific fields using the 18m citation window. Coefficients are obtained from models including year fixed effects with standard errors clustered at the author level. The association is positive in all fields, with magnitudes ranging from 0.112 in Chemistry to 0.647 in Earth and Planetary Sciences. Intermediate effects are observed in fields such as Medicine (β = 0.373), Neuroscience (β = 0.376), and Economics, Econometrics and Finance (β = 0.388). Larger effects are found in Mathematics (β = 0.505), Computer Science (β = 0.485), and Decision Sciences (β = 0.526). The vertical dashed line denotes zero. The uniformly positive coefficients indicate that the association between AI-assisted writing and disruption is pervasive across disciplines, while substantial variation in magnitude suggests heterogeneity in how AI interacts with field-specific knowledge structures.

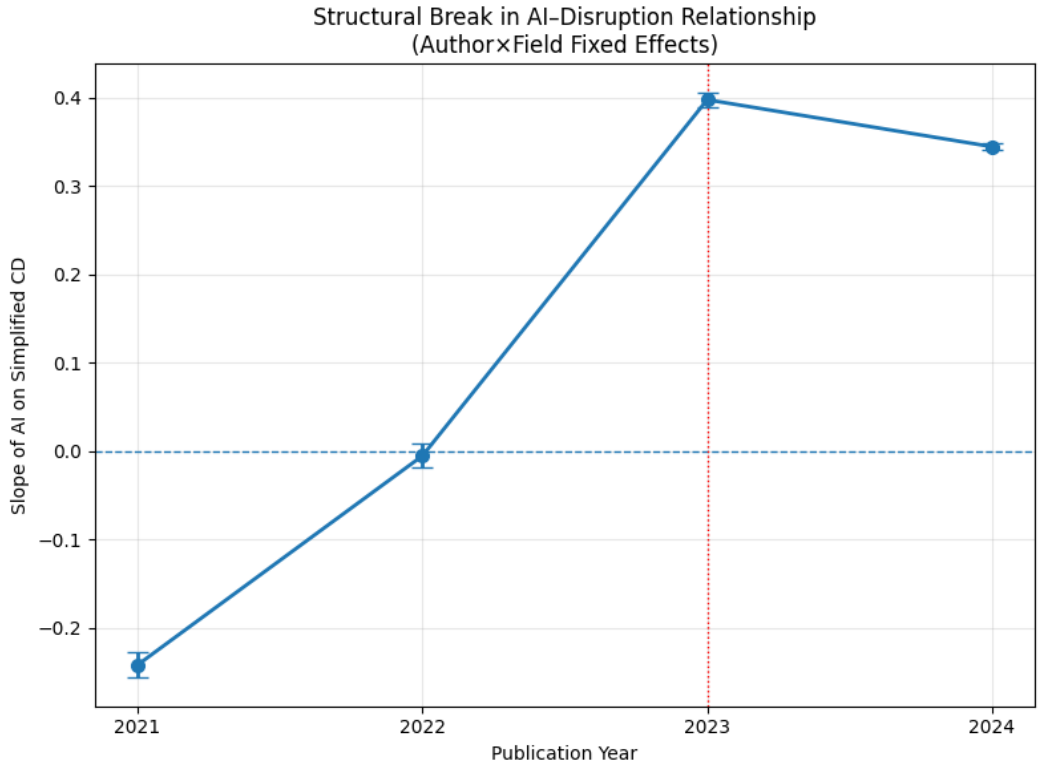

**Fig. S4. Structural break in the AI–disruption relationship under the ≥5 forward-citation specification.** Year-specific marginal effects of AI-assisted writing intensity on the simplified CD index, estimated from author–field–year panel models with author×field fixed effects using the alternative design that restricts focal works to those with at least five forward citations and uses all available forward citations rather than fixed windows. Error bars represent 95% confidence intervals. The relationship is negative in 2021 ($\beta_{(2021)}$ = −0.242, SE = 0.007), indistinguishable from zero in 2022 ($\beta_{(2022)}$ = −0.006, SE = 0.007), and strongly positive in 2023 ($\beta_{(2023)}$ = 0.397, SE = 0.004) and 2024 ($\beta_{(2024)}$ = 0.344, SE = 0.002), confirming that the main reversal is not driven by the fixed-window design.

**Table S1. Structural change in the AI–disruption relationship across citation windows.** Coefficients represent year-specific marginal effects of AI-assisted writing intensity on the consolidation–disruption (CD) index, estimated from author–field–year panel models with author×field and year fixed effects. Standard errors are heteroskedasticity-robust (HC1). Across all citation windows, the AI–disruption relationship shifts from negative or null in 2021–2022 to strongly positive in 2023–2024.

| Year | 12m (β, SE) | 18m (β, SE) | 24m (β, SE) |
|---|---|---|---|
| 2021 | −0.227 (0.011) | −0.165 (0.009) | −0.095 (0.009) |
| 2022 | −0.049 (0.010) | −0.002 (0.009) | 0.074 (0.008) |
| 2023 | 0.243 (0.007) | 0.274 (0.006) | 0.340 (0.006) |
| 2024 | 0.398 (0.004) | 0.335 (0.004) | 0.329 (0.005) |

**Table S2. Within-author acceleration of disruption associated with changes in AI-assisted writing intensity.** Coefficients are estimated from first-difference models relating changes in disruption to contemporaneous changes in AI-assisted writing intensity within the same author–field unit. All specifications include year fixed effects and heteroskedasticity-robust standard errors. The consistency across citation windows indicates that increases in AI-assisted writing intensity are associated with increases in disruption within authors.

| Window | β (ΔAI) | SE | z |
|---|---|---|---|
| 12-month | 0.152 | (0.005) | 30.9 |
| 18-month | 0.151 | (0.004) | 33.8 |
| 24-month | 0.171 | (0.005) | 34.0 |

**Table S3. Decline in cross-field recombination associated with generative AI.** Coefficients represent year-specific marginal effects of AI-assisted writing intensity on citation entropy. Prior to 2023, AI-assisted writing is associated with broader cross-field recombination. Following the diffusion of generative AI, this relationship weakens substantially, indicating reduced citation breadth.

| Year | 12m (β, SE) | 18m (β, SE) | 24m (β, SE) |
|---|---|---|---|
| 2021 | 1.251 (0.005) | 1.277 (0.005) | 1.287 (0.005) |
| 2022 | 1.200 (0.007) | 1.225 (0.007) | 1.238 (0.006) |
| 2023 | 0.721 (0.006) | 0.739 (0.006) | 0.748 (0.006) |
| 2024 | 0.437 (0.006) | 0.455 (0.006) | 0.484 (0.006) |

**Table S4. Decline in citation concentration (HHI) associated with generative AI.** Coefficients represent year-specific marginal effects of AI-assisted writing intensity on citation concentration (HHI). Prior to 2023, AI-assisted writing is associated with lower concentration. After 2023, this relationship weakens substantially, indicating a shift toward more citation concentration.

| Year | 12m (β, SE) | 18m (β, SE) | 24m (β, SE) |
|---|---|---|---|
| 2021 | −0.549 (0.002) | −0.561 (0.002) | −0.567 (0.002) |
| 2022 | −0.527 (0.003) | −0.539 (0.003) | −0.545 (0.003) |
| 2023 | −0.320 (0.003) | −0.328 (0.003) | −0.333 (0.003) |
| 2024 | −0.198 (0.003) | −0.207 (0.002) | −0.222 (0.003) |

**Table S5. Robustness of AI–disruption estimates to alternative variance estimators.** Comparison of coefficient estimates and standard errors from models using heteroskedasticity-robust (HC1) and clustered standard errors. The table reports coefficients from the fully residualized two-way fixed effects specification, with standard errors clustered at the author–field level. Coefficient estimates are identical across specifications, while clustered standard errors are modestly larger, reflecting within-panel correlation. The statistical significance and substantive interpretation of the results remain unchanged, indicating that inference is robust to the choice of variance estimator.

| Variable | Coef | SE (HC1) | SE (Clustered) |
|---|---|---|---|
| 2021 | -0.165 | 0.007 | 0.009 |
| 2022 | -0.003 | 0.007 | 0.009 |
| 2023 | 0.274 | 0.005 | 0.006 |
| 2024 | 0.335 | 0.003 | 0.004 |

**Table S6. Robustness of the AI–disruption relationship to controlling for team size.** Coefficients are estimated from a two-way fixed effects model implemented via within-transformation. Team size is measured as the average number of authors per paper within each author–field–year cell. Standard errors are clustered at the author–field level. The AI coefficients correspond to year-specific marginal effects.

| Variable | Coefficient | SE | z | p-value |
|---|---|---|---|---|
| AI × 2021 | −0.172 | 0.009 | −18.15 | <0.001 |
| AI × 2022 | −0.007 | 0.009 | −0.76 | 0.448 |
| AI × 2023 | 0.273 | 0.006 | 46.64 | <0.001 |
| AI × 2024 | 0.335 | 0.004 | 92.39 | <0.001 |
| Team size | −0.003 | 0.000 | −6.49 | <0.001 |

# Supplementary Methods

## S1. Data construction and panel design
### S1.1 Source data and sample construction

All bibliometric data are derived from the OpenAlex Walden snapshot, including works, authorships, references, forward citations, and topic classifications(22). The focal dataset consists of full-text research articles from PubMed Central published between 2021 and 2024 for which AI-assisted writing intensity can be measured(13). The initial corpus includes approximately two million focal works.

Each work is linked to its authors, primary field classification, reference list, and forward citation network. To support within-author identification, we aggregate paper-level outcomes to the author–field–year level while only keeping author with more than one author×year observation for the event model. The resulting panel includes:
- 5,437,536 observations (12-month window)
- 5,634,835 observations (18-month window)
- 5,124,693 observations (24-month window)

All panels cover four publication years (2021–2024). Aggregation reduces idiosyncratic noise from individual papers and enables fixed-effects estimation using within-author variation over time.

Our analysis is based on a large corpus of full-text articles from PubMed Central, which is enriched in biomedical research and may not be fully representative of all scientific output. However, the identification strategy relies on within-author variation over time, which differences out time-invariant characteristics of authors and fields. As a result, systematic differences between PMC and non-PMC papers do not bias the estimated relationships. Moreover, the sample spans a broad range of disciplines, and the observed patterns are consistent across fields. Finally, the mechanisms examined are structural features of scientific production that are not specific to any single domain. These considerations suggest that the results capture general changes in how research is organized, rather than artifacts of the PMC sample.

## S2. Measurement of AI-assisted writing intensity

We model each document $d$ as a mixture of human- and AI-generated sentence distributions and estimate the mixing parameter $\lambda_d$ via maximum likelihood, where sentence likelihoods are computed from token-level occurrence probabilities under reference human ($P$) and AI ($Q$) corpora (13). This distribution-based estimator exploits full lexical probability structure rather than sparse stylistic markers, distinguishing it from keyword- or classifier-based approaches.

A key identification concern is confounding from secular shifts in academic writing style. The estimator mitigates this through two properties. First, inference depends on likelihood ratios $Q(x_i)/P(x_i)$ aggregated over sentences, which require systematic alignment with the AI distribution across the vocabulary, not isolated lexical changes. Second, $P$ and $Q$ are fixed, population-level reference distributions, so gradual stylistic drift that does not approximate Q will not produce large increases in $\lambda_d$. Consistent with this identification strategy, estimated AI involvement exhibits a structural break following the release of ChatGPT, and strong cross-sectional heterogeneity by linguistic environment and author characteristics, patterns difficult to reconcile with uniform stylistic convergence. Prior validation shows low estimation error (<3.5%) and agreement with independent detection signals.

### S3. Construction of the disruption index
### S3.1 Simplified consolidation–disruption index
The primary outcome in the main text is the simplified consolidation–disruption (CD) index(12, 16):

$$CD_{simplified} = \frac{N_i - N_j}{N_i + N_j}$$

where:
- $N_i$ counts forward citations to the focal work that do not cite any of its references;
- $N_j$ counts forward citations that co-cite both the focal work and at least one of its references.

The simplified formulation focuses on citation displacement within the focal work's forward citation neighborhood. Positive values indicate that subsequent research builds on the focal contribution independently of its intellectual lineage.

### S3.2 Citation breadth (entropy)
To examine mechanisms, we construct two complementary measures of knowledge recombination based on reference fields. First, we compute citation breadth (entropy) as(17):

$$H = -\sum_f p_f \log(p_f)$$

where $p_f$ denotes the share of references in field *f* (n = 26). Higher values indicate broader recombination.

### S3.3 Citation concentration (HHI)
Citation concentration is measured using the Herfindahl index (HHI)(18):

$$HHI = \sum_f p_f^2$$

Higher entropy indicates broader cross-field recombination, whereas higher HHI indicates greater concentration of citations within a smaller set of fields.

### S4. Citation windows
The main analysis uses three forward citation windows: 12, 18, and 24 months, defined at the level of individual papers based on publication dates. For each focal work published at time t, only citing works published within the corresponding forward window are included. The OpenAlex version we used was updated on March 6, 2026. Therefore, for the 18-month window, only publications prior to the September 6, 2024 publication date were included; for the 24-month window, only publications prior to March 6, 2024 were included.

Using fixed forward citation windows ensures comparability across publication cohorts. Because more recent papers have had less time to accumulate citations, unrestricted citation counts would introduce truncation bias. By restricting citations to equal exposure windows, we ensure that results are not driven by differences in citation maturity. Papers that do not have a complete observation window are excluded from the relevant specification.

### S5. Fixed-effects identification

The central empirical challenge is distinguishing structural transformation from compositional change. Scholars differ systematically in baseline disruptiveness, disciplinary location, and career trajectory. A cross-sectional association between AI-assisted writing intensity and disruption (or entropy, HHI) would therefore be insufficient.

To address this, we employ fixed-effects panel models that absorb time-invariant heterogeneity(21). For the main structural break analysis, we estimate:

$$Y_{iaft} = \sum_{t} \beta_t (AI_{iaft} \cdot I_t) + \delta_{a \times f} + \lambda_t + \varepsilon_{iaft}$$

where:
- where $Y$ represents disruption, entropy, or HHI.
- $\delta_{a \times f}$ are author×field fixed effects;
- $\lambda_t$ are year fixed effects.

This specification compares the same author working within the same field across years. All identification therefore derives from within-author changes in AI-assisted writing intensity over time. Estimation is implemented via Frisch–Waugh–Lovell (FWL) residualization (20), yielding a two-way fixed effects specification in which both the dependent variable and regressors are orthogonalized with respect to the author × field and year effects prior to coefficient estimation.

### S6. Within-author behavioral change

To isolate dynamic behavioral change, we estimate models relating disruption (or entropy, HHI) to within-author changes in AI-assisted writing intensity. The acceleration specification tests whether increases in AI-assisted writing intensity predict contemporaneous increases in the outcomes, independent of stable author characteristics (21).

$$\Delta Y_{iaft} = \beta \Delta AI_{iaft} + \lambda_t + \varepsilon_{iaft}$$

### S7. Academic age and heterogeneity

Academic age is defined as years since first observed publication and capped at 50 years to address metadata outliers. After exclusion, the age distribution is smooth with mean 21.38 years. Interaction models show modest heterogeneity that is present in short citation windows but attenuates at longer windows (Figure S1).

### S8. Field-level heterogeneity

To examine whether the association between AI-assisted writing intensity and disruption varies across fields, we estimate a heterogeneity specification that interacts AI-assisted writing intensity with a field-level measure of knowledge structure. We use the data set with 18-month citation window.

We construct a baseline measure of field-level citation breadth using data from the pre-diffusion period (2021–2022) (Figures S2-S3). For each field, we compute the average citation entropy across all papers published during this period. This measure captures the diversity of knowledge inputs within a field prior to the widespread diffusion of generative AI. Higher values indicate that research in the field draws on a broader set of domains.

We then estimate the following model:
$$CD_{iaft} = \beta_1 AI_{iaft} + \beta_2 H_f + \beta_3 (AI_{iaft} \cdot H_f) + \lambda_t + \varepsilon_{iaft}$$

where $CD_{iaft}$ is the disruption index for author *a*, field *f*, and year *t*; $AI_{iaft}$ is AI-assisted writing intensity; $H_f$ is baseline field-level entropy; and $\lambda_t$ are year fixed effects.

The interaction term $\beta_3$ captures whether the relationship between AI-assisted writing intensity and disruption varies systematically with the underlying structure of knowledge in a field. A positive coefficient indicates that the association between AI and disruption is stronger in fields with more diverse knowledge inputs.

Because $H_f$ is time-invariant at the field level, this specification captures cross-field heterogeneity rather than within-author variation. It is interpreted as descriptive evidence of how the AI–disruption relationship differs across epistemic environments, rather than as a causal estimate. Standard errors are clustered at the author level to account for within-author correlation over time.

### S9. Heterogeneity by country of affiliation

To examine whether the relationship between AI-assisted writing intensity and disruption varies across linguistic contexts, we classify papers based on the country of affiliation of the first author. We use the data set with 18-month citation window. We consider two definitions of English-speaking countries.

The first is a broad definition, which includes countries in which English is spoken by at least 20% of the population or is widely used in academic and institutional settings(13). The second is a strict definition, which includes only countries where English is the dominant native language (United States, United Kingdom, Canada, Australia, New Zealand, Ireland, Singapore, and South Africa).

Using the broad definition, the interaction between AI-assisted writing intensity and English-speaking affiliation is positive but small in the pre-period ($\beta_{(2021)}$ = 0.110, SE = 0.037, p = 0.003), indicating slightly stronger associations in English-speaking contexts prior to the diffusion of generative AI. However, this interaction becomes small and statistically insignificant in the post-2023 period ($\beta_{(2023)}$ = 0.041, SE = 0.030, p = 0.175; $\beta_{(2024)}$ = 0.012, SE = 0.025, p = 0.642).

Using the strict definition yields a similar pattern. The interaction is positive and statistically significant in 2021 ($\beta_{(2021)}$ = 0.125, SE = 0.041, p = 0.002), but becomes small and statistically insignificant thereafter ($\beta_{(2023)}$ = 0.015, SE = 0.036, p = 0.672; $\beta_{(2024)}$ = 0.050, SE = 0.031, p = 0.106). In both specifications, the core AI coefficients remain unchanged, with a negative or null association prior to 2023 and a strong positive association in 2023–2024.

Across both definitions, the absence of statistically significant differences in the post-2023 period indicates that the structural shift in the AI–disruption relationship is not driven by linguistic or geographic differences. Instead, the effect appears broadly consistent across country contexts. The presence of modest differences in the pre-period suggests that pre-existing variation across linguistic environments does not persist following the widespread adoption of generative AI.

### S10. Robustness: ≥5 forward citation threshold

As a robustness check, we reproduce the earlier specification that restricts the sample to focal works with at least five forward citations and computes disruption using all available forward citations rather than fixed windows. This specification yields 6,169,966 author–field–year observations derived from approximately two million focal papers.

Under this alternative design, the temporal pattern is nearly identical to the main analysis (Figure S4). The year-specific association between AI-assisted writing intensity and disruption is negative in 2021 ($\beta_{(2021)}$ = −0.242, SE = 0.007), near zero in 2022 ($\beta_{(2022)}$ = −0.006, SE = 0.007), and strongly positive in 2023 ($\beta_{(2023)}$ = 0.397, SE = 0.004) and 2024 ($\beta_{(2024)}$ = 0.344, SE = 0.002). The implied shift between 2021 and 2023 is approximately +0.64. Because the specification still compares the same author within the same field over time, this robustness check supports the conclusion that the observed reversal is not an artifact of the fixed-window design. The implied shift between 2021 and 2023 is approximately +0.64, indicating a large and precisely estimated change.

Because this specification compares the same author within the same field over time, the observed break reflects behavioral change rather than compositional differences. The timing coincides with the diffusion of large language models following their public release in late 2022.

### S11. Full CD index (robustness)

The full CD index incorporates an additional denominator term(10, 11):

$$CD_{full} = \frac{N_i - N_j}{N_i + N_j + N_k}$$

where $N_k$ counts forward citations to referenced works that do not cite the focal paper.

We chose the simplified version because it is mathematically clearer and by not including $N_k$ in the denominator, the CD index has a more even distribution within -1 and 1. The simplified and full indices are highly correlated (r ≈ 0.95) based on a robustness test on data described in S10, and all substantive conclusions remain unchanged.

### S12. Control for collaboration structure

We use the data set with 18-month citation window. Including team size as a control does not materially alter the estimated relationship between AI-assisted writing intensity and disruption (Table S6). The magnitude, sign, and statistical significance of the AI coefficients remain virtually unchanged relative to the baseline specification.

The coefficient on team size is negative and statistically significant, indicating that, within the same author–field context, work produced by larger teams tends to be less disruptive on average. This is consistent with prior findings that larger collaborations are more likely to produce incremental contributions.

The stability of the AI coefficients after controlling for team size indicates that the observed relationship between AI-assisted writing and disruption is not driven by differences in collaboration structure. Rather, the results reflect changes in how knowledge is organized and cited, independent of team size.

We do not include controls for reference list length or forward citation counts. Reference list length is directly embedded in the construction of the consolidation–disruption (CD) index, which depends on whether citing papers co-cite a focal work and its references. Controlling for reference characteristics would therefore mechanically absorb part of the outcome and constitute over-control. Similarly, forward citation counts are closely related to the citation dynamics underlying the CD index and would partially condition on the outcome itself. Including these variables would risk removing meaningful variation associated with the mechanisms of interest.

### S13. Placebo test for pre-existing trends

To assess whether the observed structural change in the relationship between AI-assisted writing intensity and disruption reflects pre-existing trends, we conduct a placebo test using only the pre-diffusion period (2021–2022). We use the data set with 18-month citation window. If the main results were driven by gradual changes rather than a discrete shift, we would expect to observe a similar change within this earlier period.

We estimate the same fixed-effects specification as in the main analysis, restricting the sample to observations from 2021 and 2022:

$$CD_{iaft} = \beta_1 AI_{iaft} + \beta_2 (AI_{iaft} \cdot I_{2022}) + \delta_{a \times f} + \lambda_t + \varepsilon_{iaft}$$

where $I_{2022}$ is an indicator for the year 2022, $\delta_{a \times f}$ are author×field fixed effects, and $\lambda_t$ are year fixed effects. This specification tests whether the slope of the relationship between AI-assisted writing intensity and disruption changes between 2021 and 2022.

Because identification is based on within-author variation, this test isolates changes in behavior within the same individual prior to the diffusion of generative AI. A statistically insignificant interaction term indicates the absence of a structural shift in the pre-period.

The estimated interaction between AI-assisted writing intensity and the 2022 indicator is small and statistically indistinguishable from zero ($\beta_{(2022)}$ = 0.0011, SE = 0.010, p = 0.91), indicating no evidence of a pre-existing trend. This result supports the interpretation that the structural reversal observed in the main analysis is specific to the period beginning in 2023 and is not driven by gradual changes in the pre-period.

### S14. First-author robustness

Restricting the analysis to first-author observations yields substantially larger coefficient magnitudes while preserving the same temporal pattern. The relationship between AI-assisted writing intensity and disruption remains negative in 2021 ($\beta_{(2021)}$ = −0.395, SE = 0.024), attenuates in 2022 ($\beta_{(2022)}$ = −0.116, SE = 0.022), and becomes strongly positive in 2023–2024 ($\beta$ = 0.321–0.408, SE = 0.014–0.008).

The increase in magnitude relative to the full sample is consistent with measurement attenuation in the baseline specification, where AI exposure is assigned to all coauthors regardless of their role in manuscript preparation. Restricting the sample to first authors who are most directly involved in writing provides a more precise measure of AI-assisted writing intensity and yields stronger effects.

The consistency of the temporal pattern across specifications indicates that the main findings are not driven by authorship assignment and are robust to focusing on authors most directly responsible for manuscript preparation.

### S15. Clustering and inference

All models use heteroskedasticity-robust (HC1) standard errors. Fixed effects are implemented via within-transformation.

As a robustness test, we re-estimate the main model (citation window = 18 months) clustering standard errors at the author–field level to account for within-panel correlation. As shown in Table S5, coefficient estimates are identical to the baseline specification, and statistical significance remains unchanged, with clustered standard errors modestly larger than heteroskedasticity-robust estimates. These results indicate that inference is not sensitive to the choice of variance estimator.

All large-scale citation graph computations were performed in BigQuery (23) using partitioned tables restricted to the focal sample. Forward citation counts and CD components were precomputed at the paper level before aggregation.

Panel residualization and regression estimation were conducted in Python using Statsmodels OLS with robust covariance estimators.